\documentclass[submission,copyright,creativecommons]{eptcs}
\providecommand{\event}{WORDS 2011} 
\usepackage{breakurl}             

\usepackage{amssymb,amsfonts, latexsym}
\usepackage{verbatim}

\newcommand{\N}{\mathbb N}

\newcommand{\alp}{{\rm alph}}
\newcommand{\h}{\mathtt H}
\newcommand{\g}{\mathtt G}

\newcounter{dep}
\newenvironment{luet}
{
\setlength{\parskip}{5pt}
\begin{list}
{{\bf \arabic{dep}.}}{\usecounter{dep}\setlength{\topsep}{1pt}\setlength{\rightmargin}{\leftmargin}\setlength{\parsep}{3pt}\setlength{\itemsep}{2pt}}
}
{\end{list}}

\newenvironment{yks}
{
\setlength{\parskip}{1pt}
\begin{list}
{}{\usecounter{dol}\setlength{\topsep}{0pt}\setlength{\rightmargin}{\leftmargin}\setlength{\parsep}{0pt}\setlength{\itemsep}{0pt}}
}
{\end{list}}

\newtheorem{theorem}{Theorem}
\newtheorem{remark}{Rermark}
\newtheorem{corollary}{Corollary}

\title{Combinatorics on words in information security: Unavoidable regularities
 in the construction of multicollision attacks on iterated hash functions}

\author{Juha Kortelainen
\institute{Department of Information Processing Science, University of Oulu, Finland}
\email{jkortela@tols16.oulu.fi}}

\def\titlerunning{Combinatorics on words in information security}
\def\authorrunning{J. Kortelainen}

\begin{document}
\maketitle

\begin{abstract}
Classically in combinatorics on words one studies unavoidable regularities that appear in sufficiently 
long strings of symbols over a fixed size alphabet. In this paper we take another viewpoint and focus on 
combinatorial properties of long words in which the number of occurrences of any symbol is restritced by 
a fixed constant. We then demonstrate the connection  of these properties to constructing multicollision
attacks on so called generalized iterated hash functions.
\end{abstract}

\section{Introduction}

In combinatorics on words, the theory of 'unavoidable regularities' usually concerns properties of long words over a fixed finite alphabet. Famous classical results in general combinatorics and algebra such as theorems of Ramsey, Shirshov and van der Waerden can then be straightforwardly exploited (\cite{Har}, \cite{dLV}, \cite{Res}, \cite{RR1}, \cite{RR2}). The theory can be applied in the study of finiteness conditions for semigroups and (through the concept of syntactic monoid) also in regular languages and finite automata. 
To give the reader a view of the traditional basic results in unavoidable regularities we list some of its most noteworthy achievements.

Ramsey's Theorem immediately implies 

\begin{theorem}[Repeated Patterns \cite{Har}] For all positive integers $m$ and $n$ there exists a positive integer $R(m,n)$ satisfying the following. Given an alphabet $A$ and a partition $\{A_i\}_{i=1}^m$ of $A^+$ into $m$ sets, if $w\in A^+$ is any word of length at least $R(m,n)$, then $w$ is in $A^\ast A_j^nA^\ast$ for some $j\in\{1,2,\ldots,m\}$.
\end{theorem}

Let $A$ be an alphabet totally ordered by $<$. We extend the order $<$ to the \textit{lexiographic order} $<_{lex}$ of $A^\ast$ as follows. For all $u,v\in A^\ast$: $u<_{lex} v$ if either $v\in uA^+$ or $u=xay$ and $v=xbz$ for some $x,y,z\in A^\ast$ and $a,b\in A$ for which $a<b$.

Given a positive integer $n$, the word $w\in A^\ast$ is \textit{$n$-divided} if there exist words $u,x_1,x_2,\ldots, x_n,v$ in $A^\ast$ such that $w=ux_1x_2\cdots x_nv$ and 
\[
w<_{lex} ux_{\sigma(1)}x_{\sigma(2)}\cdots x_{\sigma(n)}v
\]
for any nontrivial permutation $\sigma:\,\{1,2,\ldots,n\}\rightarrow \{1,2,\ldots,n\}$. 

\begin{theorem}[Shirshov \cite{Lot,dLV,RR1}] Let $A$ be an alphabet of $k$ symbols and $p$ and $n$ positive integers with $p\ge 2n$. There then  exists a positive integer $S(k,p,n)$ such that any word in $A^\ast$ of length at least $S(k,p,n)$ either is $n$-divided or contains a $p$th power of a nonempty word of length at most $n-1$.
\end{theorem}

Let $w=a_1a_2\cdots a_m$ where $a_i\in A$ for $i=1,2,\ldots,m$. A \textit{cadence} of $w$ is any sequence $(i_1,i_2,\ldots,i_s)$ of integers such that
\[
0<i_1<i_2<\cdots <i_s\ \ \textrm{and}\ \ a_{i_1}=a_{i_2}=\cdots =a_{i_s}\ .
\]
Here the number $s$ is the \textit{order} of the cadence. The cadence $(i_1,i_2,\ldots,i_s)$ is \textit{arithmetic} if there exists a positive integer $d$ such that $i_j=i_1+(j-1)d$ for $j=1,2,\ldots,s$.

The celebrated van der Waerden's theorem can be reformulated in words as follows.

\begin{theorem}[van der Waerden \cite{Lot,dLV}] Let $A$ be an alphabet of $k$ symbols and $s$ a positive integer. There then exists a positive integer $W(k,s)$ such that any word in $A^\ast$ of length at least $W(k,s)$ possesses an arithmetic cadence of order $s$.
\end{theorem}

Combinatorial problems are also encountered in information security, for example, when designing and investigating hash functions, techniques used in message authentication and digital signature schemes. A \textit{hash function of length $n$} $($where $n\in\N_+$$)$ is a mapping $\h:\ \{0,1\}^\ast\rightarrow \{0,1\}^n$. For computing resource reasons, practical hash functions are often \textit{iterative}, i.e., they are based on some finite compression function and an initial hash value. For more details, see subsection \ref{bc2}.

An ideal hash function $\h:\, \{0,1\}^\ast\rightarrow \{0,1\}^n$ is a \textit{$($variable input length$)$ random oracle}: for
each $x\in \{0,1\}^\ast$, the value $\h(x)\in \{0,1\}^n$ is chosen uniformly at random. 

There are three main security properties that usually are required from a hash function $\h$: \textit{collision resistance}, \textit{preimage resistance}, and \textit{second preimage resistance}.

\textbf{Collision resistance:} It is computationally infeasible to find $x,x'\in\{0,1\}^\ast$, $x\neq x'$, such that $\h(x)=\h(x')$.

\textbf{Preimage resistance:} Given any $y\in\{0,1\}^n$, it is computationally infeasible to find $x\in\{0,1\}^\ast$ such that $\h(x)=y$.

\textbf{Second preimage resistance:} Given any $x\in\{0,1\}^\ast$, it is computationally infeasible to find $x'\in\{0,1\}^\ast$, $x\neq x'$, such that $\h(x)=\h(x')$.

If we want to consider the resistance properties mathematically, the concept 'computationally infeasible' should be rigorously defined. Then the security of $H$ is compared to the security of a random oracle. 

We thus say that $H$ is collision resistant (or possesses the collision resistance property) if to find $x,x'\in\{0,1\}^\ast$, $x\neq x'$, such that $\h(x)=\h(x')$ is (approximately) as difficult as to find $z,z'\in\{0,1\}^\ast$, $z\neq z'$, such that $\g(z)=\g'(z')$ for any random oracle hash function $\g$ of length $n$.

The concepts of preimage resistance and second preimage resistance can be defined analogously. 

Given a set $C\subseteq\{0,1\}^\ast$ of finite cardinality $k>1$, we say that $C$ is an \textit{$k$-collision on $\h$} if $\h(x)=\h(x')$ for all $x,x'\in C$.  Any $2$-collison is also called a collision (on $\h$). 

The sharpened definitions allow us to define a fourth security property, the so called multicollision resistance: The hash function $\h$ is \textit{multicollision resistant} if, for each $k\in\N_+$, to find an $k$-collison on $\h$ is (approximately) as difficult as to find an $k$-collison on any random oracle hash function $\g$ of length $n$.

Our conciderations are connected to multicollison resistance. Given a message $x=x_1x_2$ $\cdots x_l$ where $x_1,x_2,\ldots,x_l$ are the (equally long) blocks of $x$, the value of a generalized iterated hash function on $x$ is based on the values of a finite compression function on the message blocks $x_1,x_2,\ldots,x_l$. A nonempty word $\alpha$ over the alphabet $\{1,2,\ldots,l\}$ may then tell us in which order and how many times each block $x_i$ is expended by the compression function when producing the value of the respective generalized iterated hash function. Since the length of messages vary, we get to consider sequences of words $\alpha_1,\alpha_2,\ldots$ in which, for each $l\in\{1,2,\ldots\}$, the word $\alpha_l\in\{1,2,\ldots,l\}^\ast$ is related to messages with $l$ blocks. Practical applications state one more limitation: given a message of any length, a fixed   block is to be consumed by the compression function only a restricted number ($q$, say) of times when computing the generalized iterated hash function value. Thus in the sequence $\alpha_1,\alpha_2,\ldots$ we assume that for each $l\in\{1,2,\ldots\}$ and $m\in\{1,2,\ldots,l\}$, the number $|\alpha_l|_m$ of occurrences of the symbol $m$ in the word $\alpha_l$ is at most $q$. 

What can be said about the general combinatorial properties of the word $\alpha_l$ when $l$ grows? More generally: which kind of unavoidable regularities appear in sufficiently long words in which the number of occurrences of any symbol is bounded by a fixed constant? 

As is easy to imagine, the regularities in the words $\alpha_l$ weaken the respective generalized iterated hash function against multicollision attacks. This topic was first studied in \cite{HoS}, see also \cite{Jou,NaS,HKK,KHK,KKH,KKV}. We shall present combinatorial results on words which imply that $q$-bounded generalized iterated hash functions are not multicollision resistant.  

We proceed in the following order. In the next section basic concepts are briefly given. In the third section we first introduce the basics of generalized iterated hash functions. The connection to combinatorics on words is then established. The fourth section contains the necessary combinatorial results. Finally, the last section contains conclusions and further research proposals.

\section{Preliminaries}

Let $\N=\{0, 1, 2, \ldots\}$ be the set of all natural numbers and $\N_+=\N\setminus
\{0\}$. For each finite set $S$, let $|S|$ be the \textit{cardinality} of $S$ that is to say, the number of elements in $S$.

Let $A$ be a finite alphabet and $\alpha\in A^+$. The length of the word $\alpha$ is denoted
by $|\alpha|$; for each $a\in A$, let $|\alpha|_a$ be the number of occurrences of the letter
$a$ in $\alpha$, and let $\alp(\alpha)$ denote the set of all letters occurring in $\alpha$ at
least once. The empty word is denoted by $\epsilon$. A permutation of $A$ is any word $\beta\in A^+$ such that $|\beta|_a=1$ for each $a\in A$.

Let $B\subseteq A$. Then the \textit{projection morphism} from $A^\ast$ into $B^\ast$, denoted by  $\Pi^A_B$ is defined by $\Pi^A_B(b)=b$ if $b\in B$ and $\Pi^A_B(b)=\epsilon$ if $b\in A\setminus B$. We write $\Pi_B$ instead of $\Pi^A_B$ when $A$ is understood. Define the word $(\alpha)_B$ as follows: $(\alpha)_B=\epsilon$ if $\pi_B(\alpha)=\epsilon$ and $(\alpha)_B=a_1a_2\cdots a_s$ if $\pi_B(\alpha)\in a_1^+a_2^+\cdots
a_s^+$, where $s\in\N_+$, $a_1,a_2,\ldots,a_s\in B$, and $a_i\neq
a_{i+1}$ for $i=1,2,\ldots,s-1$.

\section{Hash functions and collisions}

In this section we first present a  compact lead-in  to (generalized) iterated hash functions. Later we wish to point out how certain results in  combinatorics on words are interconnected to successful multicollision construction on these type of hash functions. 

\subsection{Introduction to (generalized) iterated hash functions}\label{bc2}

Let $m,n\in\N_+$ be such that $m>n$. Then $H=\{0,1\}^n$ is the set of \textit{hash values} (of length $n$) and $B=\{0,1\}^m$) is the set of \textit{message blocks} (of length $m$). Any $w\in B^+$ is a \textit{message}. Given a mapping  $f:\,H\times B\rightarrow H$, call $f$ a \textit{compression function} (of length $n$ and block size $m$).  

Define the function   $f^+:\,H\times B^+\rightarrow H$ inductively as follows. For each $h\in H$, $b\in B$ and $x\in B^+$, let $f^+(h,b)=f(h,b)$ and
$f^+(h,b\,x)=f^+(f(h,b),x)$. Note that $f^+$ is nothing but an iterative generalization of the compression function $f$.

Let $l\in\N_+$ and $\alpha$ be a nonemptyword such that $\alp(\alpha)\subseteq\N_l$. Then $\alpha=i_1i_2\cdots i_s$, where $s\in\N_+$ and $i_j\in\N_l$ for $j=1,2,\ldots,s$. Define the \emph{iterated compression function} $f_\alpha:\,H\times B^l\rightarrow H$ (based on $\alpha$
and $f$) by 
\[
f_\alpha(h,b_1b_2\cdots b_l)\,=\,f^+(h,b_{i_1}b_{i_2}\cdots b_{i_s})
\]
for each $h\in H$ and $b_1,b_2,\ldots,b_l\in B$. Note that clearly $\alpha$ only declares how many times and in which order the message blocks $b_1,b_2,\ldots,b_l$ are used when creating the (hash) value $f_{\alpha}(h,b_1b_2\ldots b_l)$ of the message $b_1b_2\cdots b_l$. 

Given $k\in\N_+$ and $h_0\in H$, a \textit{$k$-collision $($with  initial value $h_0$$)$ in the iterated compression function $f_\alpha$} is a set $C\subseteq B^l$ such that the following holds:
\begin{luet}
\item[$1.$] The cardinality of $C$ is $k$; 
\item[$2.$] For  all $u,v\in C$ we have $f_\alpha(h_0,u)\,=\,f_\alpha(h_0,v)$; and
\item[$3.$] For any pair of distinct messages $u=u_1u_2\cdots u_l$ and $v=v_1v_2\cdots v_l$ in $C$ such that $u_i,v_i\in B$ for $i=1,2,\ldots,l$, there  exists $j\in\{1,2,\ldots,l\}$ for which $u_j\neq v_j$.
\end{luet}

For each $j\in\N_+$, let now $\alpha_j\in\N_j^+$ be such that
$\alp(\alpha_j)=\N_j$. Denote $\hat{\alpha}=(\alpha_1,\alpha_2,
\ldots)$. Define the \emph{generalized iterated hash function} (a $\mathrm{gihf}$ for short)
$\h_{\hat{\alpha},f}:\, H\times B^+\rightarrow
H$ (based on $\hat{\alpha}$ and $f$) as follows: Given the
initial value $h_0\in H$ and the message $x\in B^j$, $j\in\N_+$, let
\[
\h_{\hat{\alpha},f}(h_0,x)\,=\,f_{\alpha_j}(h_0,x)\ .
\]

Thus,  given any message $x$ of $j$ blocks and hash value $h_0$, to obtain the value $\h_{\hat{\alpha},f}(h_0,x)$, we just pick the word $\alpha_j$ from the sequence $\hat{\alpha}$ and compute $f_{\alpha_j}(h_0,x)$. For more details, see \cite{KHK} and \cite{HoS}.

\begin{remark}\label{Tra}
A traditional iterated hash function $\h:\, B^+\rightarrow
H$ based on $f$ (with initial value $h_0\in H$) can of
course be defined by $\h(u)=f^+(h_0,u)$ for each
$u\in B^+$. On the other hand $\h$ is a generalized iterated
hash function $\h_{\hat{\alpha},f}:\, H\times
B^+\rightarrow H$ based on $\hat{\alpha}$ and $f$
where $\hat{\alpha}=(1,1\cdot 2,1\cdot2\cdot3,\ldots)$ and the initial
value is fixed as $h_0$. Note that almost all hash functions used nowadays in practise are of this form.  
\end{remark}

Given $k\in\N_+$ and $h_0\in H$, a \textit{k-collision in the generalized iterated hash function $\h_{\hat{\alpha},f}$ $($with initial value $h_0$$)$} is a set
$C$ of $k$ messages such that for all $u,v\in C$,
$|u|=|v|$ and $\h_{\hat{\alpha},f}(h_0,u)=\h_{\hat{\alpha},f}(h_0,v)$. Now suppose
that $C$ is a $k$-collision in $\h_{\hat{\alpha},f}$ with
initial value $h_0$. Let $l\in\N_+$ be such that
$C\subseteq B^l$, i.e., the length in blocks of each message in
$C$ is $l$. Then, by definition, for each $u,v\in C$, the equality
$f_{\alpha_l}(h_0,u)=f_{\alpha_l}(h_0,v)$ holds. Since
$\alp(\alpha_l)=\N_l$ (and thus each symbol in $\N_l$ occurs in $\alp(\alpha)$), the set $C$ is a $k$-collision
in $f_{\alpha_l}$ with initial value $h_0$. Thus, a $k$-collision in the generalized iterated hash function $\h_{\hat{\alpha},f}$ necessarily by definition, is a $k$-collision in the iterated compression function $f_{\alpha_l}$ for some $l\in\N_+$.

Now, in our security model, the \textit{attacker} tries to find a $k$-collision in $\h_{\hat{\alpha},f}$. We assume that the attacker knows how $\h_{\hat{\alpha},f}$ depends on the respective compression function $f$ (i.e., the attacker knows $\hat\alpha$), but sees $f$ only
as a black box. She/he does not know anything about the internal
structure of $f$ and can only make \textit{queries} (i.e., pairs $(h,b)\in H\times B$) on
$f$ and get the respective \textit{responses} (values $f(h,b)\in H$).

We thus define the \textit{$($message$)$ complexity of a $k$-collision} in $\h_{\hat{\alpha},f}$ to be the expected number of queries on the compression function $f$ that is needed to create a multicollision of size $k$ in $\h_{\hat{\alpha},f}$ with any initial value $h\in H$.

According to the (generalized) \textit{birthday paradox}, a $k$-collision for any compression function $f$ of length $n$ can be found 
(with probability approx. $\frac{1}{2}$) by hashing  $(k!)^{\frac{1}{k}}2^{\frac{n(k-1)}{k}}$ messages \cite{STKT} if 
we assume that there is no memory restrictions. Two remarks can be made immediately: 

\begin{itemize}
\item[$\bullet$] In the case $k=2$ approximately $\sqrt{2}\cdot 2^{\frac{n}{2}}$ hashings (queries on $f$) are needed; intuitively many of us 
would expect the number to be around $2^{n-1}$.
\item[$\bullet$] For each $k$ in $\N_+$, finding a $(k+1)$-collision consumes much more resources than finding a 
$k$-collision.
\end{itemize}

Of course, when attacking, for instance, against an iterated hash function based on a random oracle compression function of length $n$, the attacker needs a lot of computing power when $n$ is large; to create a $2$-collison requires approximately $\sqrt{2}\cdot 2^{\frac{n}{2}}$ queries on $f$  and this is resource consuming.   

The paper \cite{Jou} presents a clever way to find a $2^r$-collision in the traditional iterated hash function $\h$ (see Remark \ref{Tra}) for any $r\in\N_+$.   
The attacker starts from the initial value $h_0$ and searches two distinct message blocks $b_1$, $b'_1$ such that $f(h_0,b_1)=f(h_0,b'_1)$ and denotes  
$h_1=f(h_0,b_1)$. By the birthday paradox, the expected number of queries on $f$ is $\tilde a\,2^{\frac{n}{2}}$ , where $\tilde a$ is approximately $2.5$.
Then, for each $i=2,3,\ldots,r-1$, the attacker continues by searching message blocks $b_i$ and $b'_i$ such that $b_i\neq b'_i$ and 
$f(h_{i-1},b_i)=f(h_{i-1},b'_i)$ and and stating $h_i=f(h_{i-1},b_i)$. Now the set 
$C=\{b_1,b'_1\}\times\{b_2,b'_2\}\times \cdots \times\{b_r,b'_r\}$ is $2^r$-collision in $\h$. The expected number of queries on $f$ is clearly $\tilde a\,r 2^{\frac{n}{2}}$, i.e., the work the attacker is expected to do is only $r$ times greater than the work she or he has to do to find a single $2$-collision. The size of the multicollision grows exponentially while the need of resources increases linearly.  

The question arises whether or not the ideas of Joux can be applied in a more broad setting, i.e., can Joux's approach be used to multicollisions in certain generalized iterated hash functions? 

In the following we shall see that this indeed is possible. Call the sequence $\hat{\alpha}=(\alpha_1,\alpha_2\ldots)$
$q$\textit{-bounded}, $q\in\N_+$, if $|\alpha_j|_i\leq q$ for each $j\in\N_+$ and $i\in\N_j$. The $\mathrm{gihf}$ $\h_{\hat{\alpha},f}$ is \textit{$q$-bounded} if $\hat\alpha$ is $q$-bounded. Note that Joux's method is easy to apply to any $1$-bounded generalized iterated hash function. 

Is it possible to extend Joux's method furthermore to be adapted to $q$-bounded $\mathrm{gihf}$s, when $q>1$? This question has been investigated first for $2$-bounded $\mathrm{gihf}$s in \cite{NaS} and then for any $q$-bounded $\mathrm{gihf}$ in \cite{HoS} (see also \cite{KHK}). It turned out that it is possible to create $2^r$-collision in any $q$-bounded $\mathrm{gihf}$ with  $O(g(n,q,r)\,2^{\frac{n}{2}})$ queries on $f$, where $g(n,q,r)$ is function of $n,q$ and $r$ which is polynomial with respect to $n$ and $r$ but double exponential with respect to $q$. 

The idea behind the successful construction of the attack is the fact that since
$\hat\alpha$ is $q$-bounded, unavoidable regularities start to appear
in the word $\alpha_l$ of $\hat\alpha$ when $l$ is increased. More
accurately, choosing $l$ large enough, yet so that $|\alp(\alpha_l)|$ depends only
polynomially on $n$ and $r$ (albeit double exponentially in $q$), a number $p\in\{1,2,\ldots,q\}$ and a set $A\subseteq
\alp(\alpha_l)$ of cardinality $|A|=n^{p-1}r$ can be found such that
\begin{yks}
\item[(P1)] $\alpha_l=\beta_1\beta_2\cdots\beta_p$ the word 
  $(\beta_i)_A$ is a permutation of $A$ for $i=1,2,\ldots,p$; and
\item[(P2)] for any $i\in\{1,2,\ldots,p-1\}$, if
  $(\beta_i)_A=z_1z_2\cdots z_{n^{p-i}r}$ is a factorization of
  $(\beta_i)_A$ such that $|\alp(z_j)|=n^{i-1}$ for $j=1,2,\ldots
  n^{p-i}r$ and $(\beta_{i+1})_A=u_1u_2\cdots u_{n^{p-i+1}r}$ is a
  factorization of $(\beta_{i+1})_A$ such that $|\alp(u_j)|=n^i$
  for $j=1,2,\ldots n^{p-i+1}r$, then for each $j_1\in\{1,2,\ldots,$
  $n^{p-i}r\}$, there exists $j_2\in\{1,2,\ldots,$ $n^{p-i-1}r\}$ such
  that $\alp(z_{j_1})\subseteq \alp(u_{j_2})$.
\end{yks}

The property (P1) allows the attacker construct a $2^{|A|}$-collision $C_1$ in
$f_{\beta_1}$ with any initial value $h_0$ so that the expected number
of queries on $f$ is $\tilde a(|\beta_1|\,2^{\frac{n}{2}})$. The property (P2) ensures that based on the multicollision  guaranteed by (P1),
the attacker can proceed and, for $i=2,3,\ldots,p$, create a $2^{n^{p-i}r}$-collision $C_{i}$ in $f_{\beta_1\beta_2\cdots \beta_{i}}$ so that the expected number of queries on $f$ is $\tilde a|\beta_1\beta_2\cdots\beta_{i}|\,2^{\frac{n}{2}}$. Thus finally a $2^r$-collision of complexity $\tilde a|\alpha|\,2^{\frac{n}{2}}$ in 
$\h_{\hat{\alpha},f}$ is generated.

Finally on the basis of the previous attack construction and (the future) Theorem \ref{main2}, the following can be proved  (\cite{KKV}).

\begin{theorem}\label{main3}
Let $m,\,n$ and $q$ be positive integers such that $m > n$ and $q>1$, $f:\,\{0,1\}^n\times \{0,1\}^m \rightarrow \{0,1\}^n$ a compression function, and $\hat\alpha=(\alpha_1,\alpha_2,\ldots)$ a $q$-bounded sequence of words such that $\alp(\alpha_l)=\N_l$ for each
$l\in\N_+$. Then, for each $r\in\N_+$, there exists a $2^r$-collision attack on the generalized iterated hash function $\h_{\hat{\alpha},f}$ such that the expected number of queries on $f$ is at most $\tilde a\,q\,N(n^{(q-1)^2}r^{2q-3},q)\,2^{\frac{n}{2}}$.
\end{theorem}

\begin{remark}
The inequality $N(m,q)<m^{2^{q-1}}$ (see Theorem \ref{main1}) implies that $$N(n^{(q-1)^2}r^{2q-3},q)\,<\,n^{(q-1)^22^{q-1}}r^{(2q-3)\,2^{q-1}}\ .$$ 
\end{remark}

The results in \cite{STKT} imply that, given a random oracle hash function $\g$ of length $2^n$, the expected number of queries on $\g$ to find a $2^r$-collision is in $\Omega(2^{n\frac{2^r-1}{2^r}})$. 

Call a generalized iterated hash function bounded if it is $q$-bounded for some $q\in\N_+$. 

\begin{corollary}
There does not exist a bounded generalized iterated hash function that is multicollision resistant.
\end{corollary}

\subsection{Essential combinatorial results}

We state a list of combinatorial results that imply Theorem \ref{main3}. The main result in stated is the form of classical combinatorial theorems. For a proof, see \cite{KKV}.  

\begin{theorem} \label{main1}
For all positive integers  $m$ and $q$ there exists a (minimal) positive integer $N(m,q)$ such that if $\alpha$ is a word
for which $|\alp(\alpha)|\geq N(m,q)$ and $|\alpha|_a\le q$ for
each $a\in \alp(\alpha)$, there exist $A\subseteq \alp(\alpha)$
with $|A|=m$, and $p\in\{1,2,\ldots, q\}$, as well as words
$\alpha_1, \alpha_2, \ldots, \alpha_p$ such that $\alpha = \alpha_1
\alpha_2 \cdots \alpha_p$ and for all $i\in\{1,2, \ldots, p\}$, the word $(\alpha_i)_A$ is a permutation of $A$. Moreover, for all $m,q\in\N_+$ we have $N(m,q+1)\leq N(m^2-m+1,q)$.
\end{theorem}

\begin{remark} \label{rem1}
Let $m\in\N_+$. In the case $q=2$, the previous theorem gives us the boundary value $N(m,2)=m^2-m+1$. Let 
\[
A\,=\,\{a_{i,j}|i=1,2,\ldots,m-1,\,j=1,2,\ldots,m\}
\]
be an alphabet of $m(m-1)$ symbols. Let furthermore 
\[
\gamma_i\,=\,a_{i,1}a_{i,2}\cdots a_{i,m-1}a_{i,m}a_{i,m-1}a_{i,m-2}\cdots a_{i,1}
\]
for $i=1,2,\ldots,m-1$ and $\alpha=\gamma_1\gamma_2\cdots\gamma_{m-1}$. It is quite straightforward to see that there does not exist an $m$-letter subalphabet of $A$ such that either\, (i) $(\alpha)_A$ is a permutation of $A$ or\  \hbox{(ii) there} exists a factorization $\alpha=\alpha_1\alpha_2$ such that $(\alpha_1)_A$ and $(\alpha_2)_A$ are both permutations of $A$. Thus $N(m,2)=m^2-m+1$ for $m\in\N_+$. 
\end{remark}

Suppose now that $A$ and $\alpha = \alpha_1\alpha_2 \cdots \alpha_p$ are as in Theorem \ref{main1}, i.e., for all $i\in\{1,2, \ldots, p\}$, the word $(\alpha_i)_A$ is a permutation of $A$. To make our multicollision attack succeed, this is not yet sufficient. We need permutations  $\beta_1$, $\beta_2$, $\ldots$, $\beta_p$ of an sufficiently large alphabet $B$ such that when factoring $\beta_i=\beta_{i1}\beta_{i2}\cdots\beta_{id_i}$ into  $d_i\in\N_+$ equal length factors for $i=1,2,\ldots,p$ where $d_j$ divides $d_{j+1}$ and the following holds: for each $i\in\{1,2,\ldots,p-1\}$ and $j_1\in\{1,2,\ldots,d_i\}$ there exists $j_2\in\{1,2,\ldots,d_{i+1}\}$ such that $\alp(\beta_{ij_1})\subseteq\alp(\beta_{i+1,j_2})$. Only then we can, starting from the first permutation (and the word $\alpha_1$) roll on our attack well. Above the permutations $\beta_1,\beta_2,\ldots,\beta_p$ are induced by the words $\alpha_1,\alpha_2, \ldots, \alpha_p$, respectively, when $\alpha$ is long enough (or equivalently, the alphabet $\alp((\alpha)$ is sufficiently large). That these permutations always can be found, is verified in the following three combinatorial results.

We wish to further study the mutual structure of permutations in long words guaranteed by Theorem \ref{main1}. By increasing the length of the word $\alpha$ the permutations are forced to possess certain stronger structural properties. The motives are, besides our interest in combinatorics on words, in information security applications. The connection of the results to creating multicollisions on generalized iterated hash functions is more accurately, albeit informally, described in Section 5.

As the first step of our reasoning we need an application of the famous Hall's Matching Theorem. For the proof, see  \cite{KHK} and \cite{HoS}.

\begin{theorem}[Partition Theorem] \label{part}
Let $k\in\N_+$ and $A$ be a finite nonempty set such that $k$ divides
$|A|$. Furthermore, let $\{B_i\}_{i=1}^k$ and $\{C_j\}_{j=1}^k$ be
partitions of $A$ such that $|B_i|=|C_j|$ for $i,j=1,2,\ldots,k$. Then
for each $x\in\N_+$ such that $|A|\geq k^3\cdot x$, there exists a
bijection $\sigma: \{1,2,\ldots,k\}\rightarrow \{1,2,\ldots,k\}$ for
which $|B_i\cap C_{\sigma(i)}|\geq x$ for $i=1,2,\ldots,k$.
\end{theorem}

The next theorem is also from \cite{KHK}. It is an inductive generalization of Partition Theorem to different size of factorizations. For the proof, see \cite{KHK}.

\begin{theorem}[Factorization Theorem] \label{perint}
Let $d_0,d_1,d_2,\ldots, d_r$, where $r\in\N_+$, be positive integers
such that $d_i$ divides $d_{i-1}$ for $i=1,2,\ldots,r$, $A$ an
alphabet of cardinality $|A|=d_0d_1^2d_2^2\cdots d_r^2$, and
$w_1,w_2,\ldots,w_{r+1}$ permutations of $A$. Then there exists a
subset $B$ of $A$ of cardinality $|B|=d_0$ such that the following
conditions are satisfied.
\begin{enumerate}
\item[\upshape{(1)}] For any $i\in\{1,2,\ldots,r\}$, if
  $\pi_B(w_i)=x_1x_2\cdots x_{d_i}$ is the factorization of 
  $\pi_B(w_i)$ and $\pi_B(w_{i+1})=y_1y_2\cdots y_{d_i}$ is the
  factorization of $\pi_B(w_{i+1})$ into $d_i$ equal length
  $(=\frac{d_0}{d_i})$ blocks, then for each $j\in\{1,2,\ldots,d_i\}$,
  there exists $j'\in\{1,2,\ldots,d_i\}$ such that
  $\alp(x_j)=\alp(y_{j'})$; and
\item[\upshape{(2)}] If $w_{r+1}=u_1u_2\cdots u_{d_r}$ is the factorization $w_{r+1}$ into $d_r$ equal length
  $(=d_0d_1^2d_2^2\cdots d_{r-1}^2d_r)$ blocks, then   $\pi_B(w_{r+1})\,=\,\pi_B(u_1)\pi_B(u_2)$ $\cdots \pi_B(u_{d_r})$
  is the factorization of $\pi_B(w_{r+1})$
  into $d_r$ equal length $(=\frac{d_0}{d_r})$ blocks.
\end{enumerate}
\end{theorem}

In fact what we need in our considerations is the following 

\begin{corollary}
Let $d_0, d$ and $r$ be positive integers such that $d$ divides $d_0$, $A$ an
alphabet of cardinality $|A|=d_0d^{2r}$, and
$w_1,w_2,\ldots,w_{r+1}$ permutations of $A$. Then there exists a
subset $B$ of $A$ of cardinality $|B|=d_0$ satisfying the following. Let $p,q\in\{1,2,\ldots,r+1\}$ and $\pi_B(w_p)=x_1x_2\cdots x_{d}$ the factorization of $\pi_B(w_p)$ and $\pi_B(w_q)=y_1y_2\cdots y_{d}$ the   factorization of $\pi_B(w_q)$ into $d$ equal length $(=\frac{d_0}{d})$ blocks, then for each $i\in\{1,2,\ldots,d\}$, there exists $j\in\{1,2,\ldots,d\}$ such that $\alp(x_i)=\alp(y_j)$.
\end{corollary}

The last result of this section combines the main result of this section (Theorem \ref{main1}) to the previous combinatorial accomplishments. Theorem \ref{main2} is indispensable for the attack constrution in the end of Section \ref{bc2}.

\begin{theorem}\label{main2}
Let $\alpha$ be a word and  $k\ge 2$, $n\ge 1$, and $q\ge 2$ integers such that 
\begin{enumerate}
\item[$(1)$] $|\alp(\alpha)|\ge N(n^{(q-1)^2}k^{2q-3},q)$; and 
\item[\rm{(2)}] $|\alpha|_a\le q$ for each $a\in \alp(\alpha)$\ .
\end{enumerate}
Then there
exists $B\subseteq \alp(\alpha)$, $p\in\{1,2,\ldots,q\}$ and a
factorization $\alpha=\alpha_1\alpha_2\cdots \alpha_p$ for which
\begin{enumerate}
\item[\rm{(3)}] $|B|=n^{p-1}k$; 
\item[\rm{(4)}] $B\subseteq \alp(\alpha_i)$ and $(\alpha_i)_B$ is a permutation of $B$ for
  $i=1,2,\ldots,p$; and
\item[\rm{(5)}] For any $i\in\{1,2,\ldots,p-1\}$, if
  $(\alpha_i)_B=z_1z_2\cdots z_{n^{p-i}k}$ is the factorization of
  of $(\alpha_i)_B$ into $n^{p-i}k$ equal length $(=n^{i-1})$
  blocks and $(\alpha_{i+1})_B=u_1u_2\cdots u_{n^{p-i-1}k}$ the
  factorization of $(\alpha_{i+1})_B$ into $n^{p-i-1}$ equal length
  $(=n^i)$ blocks, then for each $j_1\in\{1,2,\ldots,n^{p-i}k\}$,
  there exists $j_2\in\{1,2,\ldots,$ $n^{p-i-1}k\}$ such that
  $\alp(z_{j_1})\subseteq \alp(u_{j_2})$.
\end{enumerate}
\end{theorem}

\section{Conclusion}

We have considered combinatorics on words from a fresh viewpoint which is induced by applications in information security. Some small steps have already been taken in the new research frame. The results have been promising; they imply more efficient attacks on generalized iterated hash functions and, from their part, confirm the fact that the iterative structure possesses certain generic security weaknesses. 

\medskip

\noindent \textit{Research Problem}.\ \ Consider Theorem \ref{main1}. The exact value of $N(m,q)$ is known only in the cases $m=1$, $q=1$ and $q=2$: Trivially $N(1,q)=1$ and $N(m,1)=m$, furthermore $N(m,2)=m^2-m+1$ (see Remark \ref{rem1}). It is probable that in general the number $N(m,q+1)$ is significantly smaller than $N(m^2-m+1,q)$. Moreover, we have not evaluated $N(m,q)$ from below at all. Find reasonable lower and upper bounds to $N(m,q)$ for $m>1,q>2$.  

\bibliographystyle{eptcs}
\bibliography{kort}

\begin{thebibliography}{10}
\providecommand{\bibitemdeclare}[2]{}
\providecommand{\urlprefix}{Available at }
\providecommand{\url}[1]{\texttt{#1}}
\providecommand{\href}[2]{\texttt{#2}}
\providecommand{\urlalt}[2]{\href{#1}{#2}}
\providecommand{\doi}[1]{doi:\urlalt{http://dx.doi.org/#1}{#1}}
\providecommand{\bibinfo}[2]{#2}

\bibitemdeclare{inproceedings}{HKK}
\bibitem{HKK}
\bibinfo{author}{K.~Halunen}, \bibinfo{author}{J.~Kortelainen} \&
  \bibinfo{author}{Kortelainen T.} (\bibinfo{year}{2009}):
  \emph{\bibinfo{title}{Multicollision Attacks on Generalized Iterated Hash
  Functions}}.
\newblock In: {\sl \bibinfo{booktitle}{Eight Australasian Information Security
  Conference (AISC2010)}}, {\sl \bibinfo{series}{Australian Computer Science
  Communications}}~\bibinfo{volume}{32}, pp. \bibinfo{pages}{85--93}.

\bibitemdeclare{book}{Har}
\bibitem{Har}
\bibinfo{author}{Michael~A. Harrison} (\bibinfo{year}{1978}):
  \emph{\bibinfo{title}{Introduction to formal language theory}}.
\newblock \bibinfo{publisher}{Addison-Wesley Publishing Co., Reading, Mass.}

\bibitemdeclare{incollection}{HoS}
\bibitem{HoS}
\bibinfo{author}{Jonathan Hoch} \& \bibinfo{author}{Adi Shamir}
  (\bibinfo{year}{2006}): \emph{\bibinfo{title}{Breaking the ICE -– Finding
  Multicollisions in Iterated Concatenated and Expanded (ICE) Hash Functions}}.
\newblock In \bibinfo{editor}{Matthew Robshaw}, editor: {\sl
  \bibinfo{booktitle}{Fast Software Encryption}}, {\sl \bibinfo{series}{Lecture
  Notes in Computer Science}} \bibinfo{volume}{4047},
  \bibinfo{publisher}{Springer Berlin / Heidelberg}, pp.
  \bibinfo{pages}{179--194}, \doi{10.1007/11799313_12}.

\bibitemdeclare{incollection}{Jou}
\bibitem{Jou}
\bibinfo{author}{Antoine Joux} (\bibinfo{year}{2004}):
  \emph{\bibinfo{title}{Multicollisions in Iterated Hash Functions. Application
  to Cascaded Constructions}}.
\newblock In \bibinfo{editor}{Matt Franklin}, editor: {\sl
  \bibinfo{booktitle}{Advances in Cryptology -- CRYPTO 2004}}, {\sl
  \bibinfo{series}{Lecture Notes in Computer Science}} \bibinfo{volume}{3152},
  \bibinfo{publisher}{Springer Berlin / Heidelberg}, pp.
  \bibinfo{pages}{99--213}, \doi{10.1007/978-3-540-28628-8_19}.

\bibitemdeclare{unpublished}{KKV}
\bibitem{KKV}
\bibinfo{author}{J.~Kortelainen}, \bibinfo{author}{T.~Kortelainen} \&
  \bibinfo{author}{A.~Vesanen} (\bibinfo{year}{2011}):
  \emph{\bibinfo{title}{Unavoidable regularities in long words with bounded
  number of symbol occurrences}}.
\newblock \bibinfo{note}{Accepted to The 17th Annual International Computing
  and Combinatorics Conference (COCOON 2011)}.

\bibitemdeclare{article}{KHK}
\bibitem{KHK}
\bibinfo{author}{Juha Kortelainen}, \bibinfo{author}{Kimmo Halunen} \&
  \bibinfo{author}{Tuomas Kortelainen} (\bibinfo{year}{2010}):
  \emph{\bibinfo{title}{Multicollision attacks and generalized iterated hash
  functions}}.
\newblock {\sl \bibinfo{journal}{J. Math. Cryptol.}}
  \bibinfo{volume}{4}(\bibinfo{number}{3}), pp. \bibinfo{pages}{239--270},
  \doi{10.1515/JMC.2010.010}.

\bibitemdeclare{incollection}{KKH}
\bibitem{KKH}
\bibinfo{author}{T.~Kortelainen}, \bibinfo{author}{J.~Kortelainen} \&
  \bibinfo{author}{J.~Halunen} (\bibinfo{year}{2010}):
  \emph{\bibinfo{title}{Variants of multicollision attacks on iterated hash
  functions}}.
\newblock In: {\sl \bibinfo{booktitle}{6th China International Conference on
  Information Security and Cryptology (Inscrypt 2010)}}, pp.
  \bibinfo{pages}{159--175}.

\bibitemdeclare{book}{Lot}
\bibitem{Lot}
\bibinfo{author}{M.~Lothaire} (\bibinfo{year}{1997}):
  \emph{\bibinfo{title}{Combinatorics on words}}.
\newblock \bibinfo{series}{Cambridge Mathematical Library},
  \bibinfo{publisher}{Cambridge University Press},
  \bibinfo{address}{Cambridge}, \doi{10.1017/CBO9780511566097}.

\bibitemdeclare{book}{dLV}
\bibitem{dLV}
\bibinfo{author}{Aldo de~Luca} \& \bibinfo{author}{Stefano Varricchio}
  (\bibinfo{year}{1999}): \emph{\bibinfo{title}{Finiteness and regularity in
  semigroups and formal languages}}.
\newblock \bibinfo{series}{Monographs in Theoretical Computer Science. An EATCS
  Series}, \bibinfo{publisher}{Springer-Verlag}, \bibinfo{address}{Berlin}.

\bibitemdeclare{article}{NaS}
\bibitem{NaS}
\bibinfo{author}{Mridul Nandi} \& \bibinfo{author}{Douglas~R. Stinson}
  (\bibinfo{year}{2007}): \emph{\bibinfo{title}{Multicollision attacks on some
  generalized sequential hash functions}}.
\newblock {\sl \bibinfo{journal}{IEEE Trans. Inform. Theory}}
  \bibinfo{volume}{53}(\bibinfo{number}{2}), pp. \bibinfo{pages}{759--767},
  \doi{10.1109/TIT.2006.889721}.

\bibitemdeclare{article}{Res}
\bibitem{Res}
\bibinfo{author}{A.~Restivo} (\bibinfo{year}{1977}): \emph{\bibinfo{title}{Mots
  sans r\'ep\'etitions et langages rationnels born\'es}}.
\newblock {\sl \bibinfo{journal}{RAIRO Informat. Th\'eor.}}
  \bibinfo{volume}{11}(\bibinfo{number}{3}), pp. \bibinfo{pages}{197--202, iv}.

\bibitemdeclare{article}{RR1}
\bibitem{RR1}
\bibinfo{author}{Antonio Restivo} \& \bibinfo{author}{Christophe Reutenauer}
  (\bibinfo{year}{1983}): \emph{\bibinfo{title}{Some applications of a theorem
  of {S}hirshov to language theory}}.
\newblock {\sl \bibinfo{journal}{Inform. and Control}}
  \bibinfo{volume}{57}(\bibinfo{number}{2-3}), pp. \bibinfo{pages}{205--213},
  \doi{10.1016/S0019-9958(83)80044-8}.

\bibitemdeclare{article}{RR2}
\bibitem{RR2}
\bibinfo{author}{Antonio Restivo} \& \bibinfo{author}{Christophe Reutenauer}
  (\bibinfo{year}{1985}): \emph{\bibinfo{title}{Rational languages and the
  {B}urnside problem}}.
\newblock {\sl \bibinfo{journal}{Theoret. Comput. Sci.}}
  \bibinfo{volume}{40}(\bibinfo{number}{1}), pp. \bibinfo{pages}{13--30},
  \doi{10.1016/0304-3975(85)90156-2}.

\bibitemdeclare{incollection}{STKT}
\bibitem{STKT}
\bibinfo{author}{Kazuhiro Suzuki}, \bibinfo{author}{Dongvu Tonien},
  \bibinfo{author}{Kaoru Kurosawa} \& \bibinfo{author}{Koji Toyota}
  (\bibinfo{year}{2006}): \emph{\bibinfo{title}{Birthday paradox for
  multi-collisions}}.
\newblock In: {\sl \bibinfo{booktitle}{Information security and
  cryptology---{ICISC} 2006}}, {\sl \bibinfo{series}{Lecture Notes in Comput.
  Sci.}} \bibinfo{volume}{4296}, \bibinfo{publisher}{Springer},
  \bibinfo{address}{Berlin}, pp. \bibinfo{pages}{29--40},
  \doi{10.1007/11927587_5}.

\end{thebibliography}

\end{document}